\documentclass[showpacs,preprintnumbers,amsmath,amssymb,prb,singlecolumn,superscriptaddress]{revtex4}

\usepackage{graphicx}   % Include figure files
\usepackage{dcolumn}    % Align table columns on decimal point
\usepackage{bm}         % bold math

\usepackage{epsfig}

\begin{document}

\title{Dirac gap-induced graphene quantum dot in an electrostatic potential}

\author{G. Giavaras}
\affiliation{Advanced Science Institute, RIKEN, Wako-shi, Saitama
351-0198, Japan}

\author{Franco Nori}
\affiliation{Advanced Science Institute, RIKEN, Wako-shi, Saitama
351-0198, Japan} \affiliation{Department of Physics, The
University of Michigan, Ann Arbor, MI 48109-1040, USA }

%\date{\today}


\begin{abstract}
A spatially modulated Dirac gap in a graphene sheet leads to
charge confinement, thus enabling a graphene quantum dot to be
formed without the application of external electric and magnetic
fields [Appl. Phys. Lett. \textbf{97}, 243106 (2010)]. This can be
achieved provided the Dirac gap has a local minimum in which the
states become localised. In this work, the physics of such a
gap-induced dot is investigated in the continuum limit by solving
the Dirac equation. It is shown that gap-induced confined states
couple to the states introduced by an electrostatic quantum well
potential. Hence the region in which the resulting hybridized
states are localised can be tuned with the potential strength, an
effect which involves Klein tunneling. The proposed quantum dot
may be used to probe quasi-relativistic effects in graphene, while
the induced confined states may be useful for graphene-based
nanostructures.
\end{abstract}

\maketitle

\section{Introduction}

Quantum dots have been studied extensively over the last years
because they allow fundamental laws of quantum physics to be
probed and they might also find applications in nanoelectronics as
detectors and transistors. Although a dot in a semiconductor
heterostructure can be routinely formed, in monolayer graphene the
situation is
different.~\cite{abergel2010,rozhkov,ponomarenko,nanoflakes} An
ideal graphene sheet is gapless, its energy dispersion is linear,
$E$=$\pm v_{\text{F}}|\bf{p}|$, and its charge carriers are
massless Dirac particles. Therefore, they exhibit Klein tunneling,
which allows massless particles to tunnel through any
electrostatic potential barrier. This property excludes the
possibility of fabricating a graphene dot simply by using
electrostatic gates, as in common semiconductors.

In particular, the states of an electrostatic graphene dot are
deconfined, i.e., they have an oscillatory tail outside the dot
region, which is a direct consequence of the Klein
tunneling.~\cite{matulis2008,chen,apalkov} This essentially means
that the electrons can only spend a finite time interval inside
the dot, and it has been shown theoretically that this time
interval is sensitive to the details of the potential profile and
the energy of the quantum state.~\cite{apalkov} A uniform magnetic
field leads to confined states (bound states) which decay
exponentially outside the dot, provided the electrostatic
potential rises slowly compared to the magnetic vector
potential.~\cite{giavaras2009,giavaras2010a} Thus a graphene
quantum dot can be formed with a uniform magnetic field. This
property has also potential applications in graphene
waveguides.~\cite{bliokh,bliokh2}

It is experimentally possible to induce an energy gap in
graphene's band structure, referred to as the Dirac
gap.~\cite{abergel2010,rozhkov} In this case the energy dispersion
becomes $E$=$\pm$$\sqrt{(v_{\text{F}}\mathbf{p})^2+\Delta^2}$,
with $2\Delta$ the value of the Dirac gap. The importance of the
gap stems from the fact that it gives rise to carriers with mass
($\sim$$\Delta$/$v^{2}_{\text{F}}$) and thus a quantum well
potential induces confinement as in common semiconductor
dots.~\cite{abergel2010,rozhkov} This happens because the Klein
tunneling is suppressed for a particle with mass, as long as its
energy lies in the gap. The quantum states with energies within
the gap decay exponentially in the barrier region, thus trapping
the electron in the dot for a theoretically infinite time.

This work considers the case of a spatially modulated Dirac gap
that enables a graphene quantum dot to be formed without the
application of external electric and magnetic
fields.~\cite{giavaras2010b} This is feasible provided the gap has
a local minimum in which the electron states become localized.
Gap-induced confined states couple with states induced by an
electrostatic quantum well potential; then this property allows
the region in which the resulting hybridized states are localised
to be tuned with the potential strength. This type of dot can be
used to probe quasi-relativistic effects in graphene, related to
Klein tunneling, while the confined states may be useful for
applications. In order to manipulate the spin states of an
electron confined in a quantum dot, magnetic fields are usually
needed. Thus, forming a graphene dot without applying additional
magnetic fields to control the charge states can be advantageous.
This is one of the motivations of our proposal. Moreover, the
density of states in a gap-induced dot can be made low enough
since only states with energies in the gap are of importance. A
magnetic dot does not have this property and it is more tricky to
induce a low density of states.~\cite{maksym2010}

The value of the Dirac gap depends on the specific experimental
technique and can range from a few to hundreds of meV. For
example, a Dirac gap has been measured in graphene grown
epitaxially on a SiC substrate.~\cite{zhou2007} A gap opening has
been demonstrated by controlling the structure of the interface
between graphene and ruthenium.~\cite{enderlein2010} Further, it
has been reported experimentally a spatially modulated and (buffer
layer) thickness-dependent Dirac gap.~\cite{vitali2008} Moreover,
it has been suggested theoretically that local strain and/or
chemical methods might also be employed to open-up and tune the
Dirac gap.~\cite{abergel2010,rozhkov,giovannetti,tiwari,ribeiro}

This paper is organized as follows. In Sec.~\ref{physmod} the
quantum dot model is presented and the necessary conditions for
confinement are derived in the presence of a magnetic vector and
scalar potentials, as well as a spatially modulated Dirac gap. In
Sec.~\ref{model} a model dot is examined in the regime where the
Dirac gap has the same spatial profile as the electrostatic
potential. In Sec.~\ref{basic} some semi-analytical results are
derived for the case of a piecewise-constant Dirac gap profile and
the general properties of the gap-induced dot are presented. The
interplay of the gap-induced dot with an electrostatic quantum
well potential is studied in Sec.~\ref{gappotential}, and
Sec.~\ref{conclusions} presents the basic conclusions of the
paper.

\section{Physical model}\label{physmod}

\subsection{Graphene quantum dot Hamiltonian}~\label{hamiltonian}

For energies near the Dirac points ($<$$\pm$ 1 eV) and in the
continuum limit, a graphene quantum dot can be described by the
effective 2$\times$2 Hamiltonian
\begin{equation}\label{dothamiltonian}
H=v_{\text{F}}\bm{\sigma}\cdot(\mathbf{p} + e\mathbf{A})
+V\mathcal{I}+\tau\Delta\sigma_{z}.
\end{equation}
The Fermi velocity $v_{\text{F}}$=$\gamma/\hbar$ is assumed to be
position-independent, where $\gamma$=646 meV nm is a band
structure parameter. Also,
$\bm{\sigma}$=$(\sigma_{x},\sigma_{y})$, $\sigma_{z}$ are the
2$\times$2 Pauli operators acting on the two carbon sublattices,
$\mathcal{I}$ is the 2$\times$2 unit matrix,
$\mathbf{p}$=$-i\hbar\bm{\nabla}$=$-i\hbar(\partial_x,\partial_y)$
is the two-dimensional momentum operator, $\mathbf{A}$ is the
vector potential that generates the magnetic field
$\mathbf{B}$=$\bm{\nabla}$$\times$$\mathbf{A}$, and $V$ is the
quantum well potential formed electrostatically, for example, due
to a gate voltage. The last term in Eq.~(\ref{dothamiltonian}),
referred to as the mass term, gives rise to an energy gap
$2\Delta$ in the spectrum of graphene, where $\tau$=+1
($\tau$=$-1$) corresponds to the $K$ ($K'$) valley. The
Hamiltonian~(\ref{dothamiltonian}) assumes no intervalley
coupling, which is a good approximation for most graphene
samples.~\cite{abergel2010,rozhkov}

For the dot model we consider that both $V$ and $\Delta$ have
cylindrical symmetry and the applied magnetic field is
perpendicular to the graphene sheet, i.e.,
$\mathbf{B}$=$B\hat{z}$. Hence only the azimuthal component
$A_{\theta}$ is nonzero, therefore
\begin{equation}
\mathbf{A}=(0,A_{\theta},0), \quad
A_{\theta}=\frac{B_{0}r^{s}}{s+1},
\end{equation}
which generates the field $B$=$B_{0}r^{s-1}$; here, $s$=1
corresponds to a uniform magnetic field. Hamiltonian
Eq.~(\ref{dothamiltonian}) is written in cylindrical coordinates
($x$=$r\cos\theta$, $y$=$r\sin\theta$) with the substitution
\begin{equation}
\partial_{x}-i\partial_{y}=e^{-i\theta}\left(\partial_{r}-\frac{i}{r}\partial_{\theta}\right),
\end{equation}
and
\begin{equation}
\bm{\sigma}\cdot\mathbf{A}=-(\sigma_{x}\sin\theta
-\sigma_{y}\cos\theta)A_{\theta}.
\end{equation}
A two-component solution $\Psi$ to the Dirac equation
\begin{equation}
H\Psi=E\Psi,
\end{equation}
with $E$ being the eigenenergy, is written in the general form
\begin{equation}
\Psi =\frac{1}{\sqrt{r}}\left(\begin{matrix}
f_{1}(r)\exp[i(m-1)\theta]
\\
i f_{2}(r)\exp(i m\theta)
\end{matrix}\right).
\end{equation}
with $m$=$0,\pm 1,$ ... being the orbital angular momentum quantum
number. The radial components $f_{1}$ and $f_{2}$ express
amplitude probabilities on the two carbon sublattices of graphene,
and they satisfy the two coupled differential equations
\begin{subequations}
\begin{eqnarray}
(V+\tau \Delta)f_{1} +\left(U+\gamma\frac{d}{dr} \right)f_{2} &=&Ef_{1},\label{radialequations1}    \\
\left(U-\gamma\frac{d}{dr}\right)f_{1}+(V-\tau \Delta)f_{2}
&=&Ef_{2}.\label{radialequations2}
\end{eqnarray}
\end{subequations}
The term $U$ is due to the angular momentum and the applied
magnetic field, and is given by
\begin{equation}\label{uterm}
U=\frac{\gamma(2m-1)}{2r}+\frac{\gamma eA_{\theta}}{\hbar}.
\end{equation}
Applying the time-reversal symmetry operator
$i\mathcal{C}\sigma_{y}$, with $\mathcal{C}$ the operator of
complex conjugation, to Eqs.~(\ref{radialequations1})
and~(\ref{radialequations2}), it can be shown that the
eigenenergies~\cite{note0} satisfy the condition
$E(m,B,\tau)$=$E(1-m,-B,-\tau)$. Also, $E(V,\tau)$=$-E(-V,-\tau)$
due to electron-hole symmetry.

\subsection{Confinement in a Dirac gap-induced dot}\label{theory}

The quantum states of a Schr\"odinger dot, formed by an
electrostatic quantum well, can be classified according to their
energy. A state is confined when its energy relative to the bottom
of the well is smaller than the well depth. This state has an
exponential tail at a large distance from the quantum well. In the
opposite regime, when the energy is larger than the well depth,
the state is deconfined and it has an oscillatory tail.

This behaviour is no longer valid for a Dirac dot formed in
graphene, because the energy spectrum of graphene is unbound and
Klein tunneling takes place. In particular, a massless Dirac
electron can tunnel through any electrostatic potential barrier,
as has been confirmed experimentally via transport measurements in
p-n junctions.~\cite{young} For this reason an electrostatic
graphene dot cannot confine electrons.

Nevertheless, confinement can be achieved in the presence of a
magnetic field and/or an energy gap in graphene's spectrum, though
for this to happen some specific conditions have to be satisfied,
which are analysed in this section. As a general rule, confined
states should decay asymptotically, i.e., at a large
($r$$\rightarrow$$\infty$) radial distance $r$, independent of
their energy, whereas deconfined states should have an oscillatory
tail. The former states are sometimes called bound and the latter
quasi-bound.

Below we employ a rigorous treatment to examine the dot
states.~\cite{giavaras2009} We derive a single second-order
differential equation for each radial component, by decoupling
Eqs.~(\ref{radialequations1}) and~(\ref{radialequations2}). Then
we identify the form of the state in the asymptotic region, i.e.,
at a large radial distance from the origin of the dot. The $f_{2}$
component satisfies
\begin{equation}\label{f2component}
\frac{d^{2}f_{2}}{dr^{2}} + a(r)\frac{df_{2}}{dr} + b(r)f_{2}=0,
\end{equation}
with the coefficients
\begin{equation}
a(r)=-\frac{V'_{+}}{V_{+}-E},\notag
\end{equation}
and
%\begin{equation}
%b(r)=a(r)\left(\frac{2m-1}{2r}+\frac{e}{\hbar}A_{\theta}\right)-\left(\frac{2m-1}{2r}+\frac{e}{\hbar}A_{\theta}\right)^{2}
%+\frac{(V_{-}-E)(V_{+}-E)}{\gamma^2}+\frac{d}{dr}\left(\frac{2m-1}{2r}+\frac{e}{\hbar}A_{\theta}\right),\notag
%\end{equation}
\begin{equation}
b(r)=a\frac{U}{\gamma}-\frac{U^2}{\gamma^2}+\frac{(V_{-}-E)(V_{+}-E)}{\gamma^2}+\frac{U'}{\gamma},\notag
\end{equation}
%\begin{equation}
%\begin{split}
%b(r)=&a(r)\left(\frac{2m-1}{2r}+\frac{e}{\hbar}A_{\theta}\right)-\left(\frac{2m-1}{2r}+\frac{e}{\hbar}A_{\theta}\right)^{2}\\
%&+\frac{(V_{-}-E)(V_{+}-E)}{\gamma^2}+\frac{d}{dr}\left(\frac{2m-1}{2r}+\frac{e}{\hbar}A_{\theta}\right),\notag
%\end{split}
%\end{equation}
with $V_{\pm}$=$V\pm\tau\Delta$ and the prime denotes
differentiation with respect to $r$. The first derivative term in
Eq.~(\ref{f2component}) complicates the analysis of the quantum
states. For this reason we eliminate it by writing $f_{2}$ in the
form $f_{2}(r)$=$g(r)u_{2}(r)$ and from Eq.~(\ref{f2component}) we
derive that
\begin{equation}
g\frac{d^{2}u_{2}}{dr^{2}} + (2g'+a g)\frac{du_{2}}{dr}+(g''+a
g'+b g)u_{2} = 0.
\end{equation}
If we choose $g$=$\exp(-\int$$a/2dr)$, the first derivative term
cancels and $u_{2}$ satisfies the Schr\"odinger-like equation
\begin{equation}\label{2ndorder}
\frac{d^{2}u_{2}}{dr^{2}} + k^2_{2}(r) u_{2} = 0,
\end{equation}
with the $r$-dependent coefficient
\begin{equation}\label{k2exact}
k^2_{2}(r)= b - \frac{1}{2}\frac{da}{dr}-\frac{a^2}{4}.
\end{equation}
Here $g$ is not an oscillatory function of $r$, therefore $f_{2}$
has the same confined or deconfined character as $u_{2}$.
Specifically Eq.~(\ref{2ndorder}) suggests that $u_{2}$ is
confined only if $k^2_{2}$ is asymptotically
($r$$\rightarrow$$\infty$) negative. Otherwise $u_{2}$ is
deconfined. The same arguments are valid for the $f_{1}$
component.~\cite{note1}

To proceed, we assume that for a large radial distance $r$ both
$V$ and $\Delta$ are constant or have a power-law dependence;
therefore in this case we derive asymptotically that
\begin{equation}\label{k2dependence}
k^2_{2}(r)\approx-\left(
\frac{eA_{\theta}}{\hbar}\right)^{2}+\left(\frac{V-E}{\gamma}\right)^{2}-\left(\frac{\tau\Delta}{\gamma}\right)^2.
\end{equation}
Equation~(\ref{k2dependence}) shows that the sign of $k^2_{2}$ is
tunable with $A_{\theta}$, $V$, $\Delta$ and energy $E$, and thus
the same is valid for the character of the dot states. Further,
the character of the states is independent of the choice of valley
$\tau$=$\pm1$, although for a fixed $m$ the energies are not the
same for both valleys.

The form of $k^{2}_{2}(r)$ in Eq.~(\ref{k2dependence}) shows that
the vector potential acts equivalently to the mass term, namely,
both have the tendency to confine the states. This can be
understood as follows. It can be seen from
Eqs.~(\ref{radialequations1}) and~(\ref{radialequations2}) that
the radial Hamiltonian which acts on $f_{1}$ and $f_{2}$ is
\begin{equation}
H_{r}=\left(%
\begin{array}{cc}
  V+\tau\Delta & U-\gamma\frac{d}{dr} \\
  U+\gamma\frac{d}{dr} & V-\tau\Delta \\
\end{array}%
\right).
\end{equation}
The similarity transformation
$P^{\dagger}H_{r}P$=$\mathcal{H}_{r}$ with the operator
\begin{equation}
P=\frac{1}{\sqrt{2}}\left(%
\begin{array}{cc}
  1 & 1 \\
 1 & -1 \\
\end{array}%
\right),
\end{equation}
results in the transformed Hamiltonian
\begin{equation}
\mathcal{H}_{r}=\left(%
\begin{array}{cc}
  V+U & \tau\Delta+\gamma\frac{d}{dr} \\
 \tau\Delta - \gamma\frac{d}{dr} & V-U \\
\end{array}%
\right).
\end{equation}
Among the terms which dominate asymptotically, $U$ and
$\tau\Delta$ are interchanged in $\mathcal{H}_{r}$ with respect to
$H_{r}$. However, asymptotically $U$$\sim$$\gamma
eA_{\theta}/\hbar$, therefore in this regime $\tau\Delta$ and
$A_{\theta}$ should act equivalently.

The necessary condition for the occurrence of confinement, i.e.,
$k^2_{2}<0$ asymptotically, means that at least one of
$A_{\theta}$, $\Delta$ has to be nonzero, otherwise $k^2_{2}>0$
leading to deconfined states. Most importantly,
Eq.~(\ref{k2dependence}) shows that confined states can be induced
even for $A_{\theta}$=0 and $V$=0 everywhere, provided that
$E^2-\Delta^2$$<$0 asymptotically. As shown in
Ref.~\onlinecite{giavaras2010b} this inequality can be satisfied
when $\Delta$ is spatially-dependent; for example, when $\Delta$
is zero within a disc area, and nonzero outside that area,
$\Delta$=$\delta_0$. Then, as shown below, discrete energy levels
in the range $|E|$$<$$\delta_0$ correspond to confined states with
a large amplitude within the disc area.

\section{A Model quantum dot system}\label{model}

To demonstrate the arguments presented in the previous section,
consider first a model quantum dot for which the electrostatic
potential $V$, and the mass term $\Delta$ have a power-law spatial
dependence. Equation~(\ref{k2dependence}) shows that if
$A_{\theta}$=0, and $V$, $\Delta$ are unequal, then confined
states occur only if ($V-\Delta$$)<$0, so that $k^{2}_{2}$$<$0. In
this case confinement is energy-independent, i.e., all the states
independent of their energy are confined. In the special limit of
$V$=$\Delta$ only energies which satisfy $-EV$$<$0 correspond to
confined states, which implies that if $V$ increases (decreases)
asymptotically then confinement occurs only for positive
(negative) energies.

The complete physical behaviour can be demonstrated when $V$ and
$\Delta$ have the same $r$-dependence, for example parabolic,
$V$=$V_{0}r^{2}$ and $\Delta$=$\Delta_{0}r^{2}$ with $V_0$,
$\Delta_0 >$ 0. Then from Eq.~(\ref{k2dependence}) the resultant
states are confined when $\Delta_{0}$$>$$V_{0}$ so that
$k^{2}_{2}$$<$0, and deconfined when $\Delta_{0}$$<$$V_{0}$ and
hence $k^{2}_{2}$$>$0. For $V_{0}$=$\Delta_{0}$ only positive
energies correspond to confined states.

To quantify this behaviour the two coupled
Eqs.~(\ref{radialequations1}) and~(\ref{radialequations2}) are
discretised on a uniform grid and the resulting matrix eigenvalue
problem is solved numerically, using a similar technique as in
Ref.~\onlinecite{giavaras2009}. The boundary conditions for the
components $f_{1}$ and $f_{2}$ which lead to a Hermitian
eigenvalue problem [Eqs.~(\ref{radialequations1})
and~(\ref{radialequations2})] are derived in
Ref.~\onlinecite{giavaras2009}, and it can be shown that these are
not modified by the presence of the mass term. So one component
has to vanish at the origin and the second has to vanish
asymptotically (at the boundary of the computational box). The
same procedure is followed in all sections.

\begin{figure}
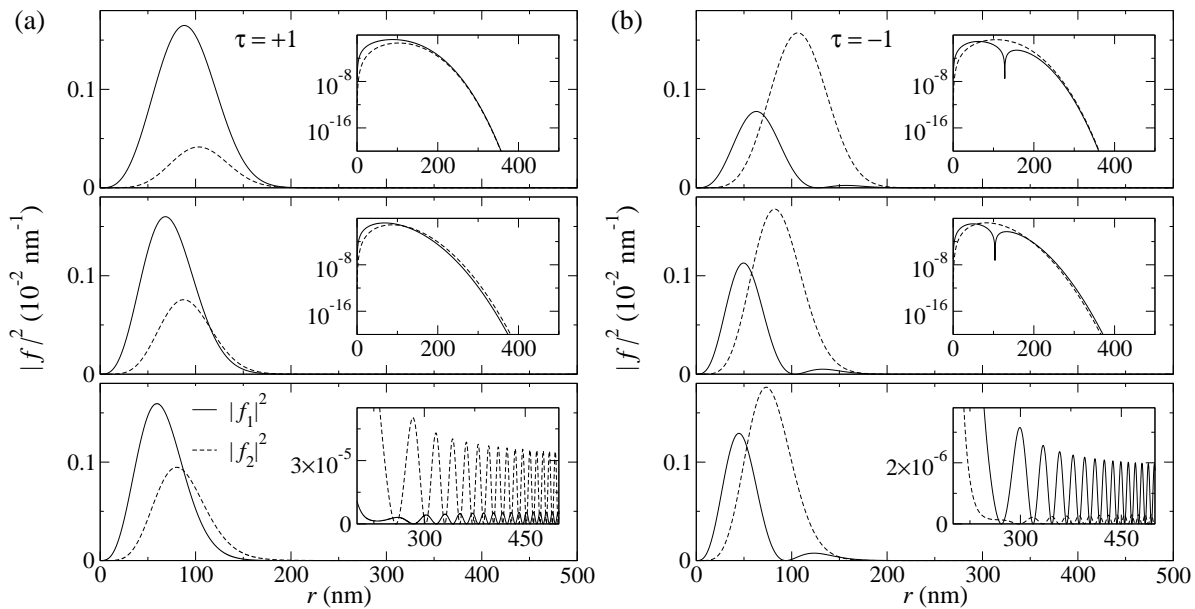

\includegraphics[height=8.cm]{fig1a_prb}
\includegraphics[height=8.cm]{fig1b_prb}
\caption{(a) Quantum states for $m$=2, $\tau$=$+$1 and different
potential strengths $V_{0}$. The mass term is modelled by
$\Delta$=$\Delta_{0}r^2$, with $\Delta_{0}$=1 $\mu$eV nm$^{-2}$,
and the electrostatic potential is modelled by $V$=$V_{0}r^{2}$.
The potential strength $V_{0}$ is from top to bottom:
$V_{0}/\Delta_{0}$=0, 0.9, 1.5. The vertical axes of the insets to
the top and middle panels are on logarithmic scale. (b) As in (a)
but for $\tau$=$-$1.}\label{qstates}
\end{figure}

Figure~\ref{qstates} shows quantum states for $\Delta_{0}$
constant, $\Delta_{0}$=1 $\mu$eV nm$^{-2}$, and three different
choices of $V_{0}$. The states plotted are those which correspond
to the lowest positive energy when $V_{0}$=0. When $V_{0}$=0
($<$$\Delta_{0}$) the states are confined and they have an
exponential tail. This demonstrates that a graphene dot can be
formed thanks to a spatially modulated Dirac gap without applying
external electric or magnetic fields. When $V_{0}$=$0.9\Delta_{0}$
($<$$\Delta_{0}$) the states are still confined though the region
within which they are localised is slightly modified by the
presence of the potential. On the other hand, when
$V_{0}$=1.5$\Delta_{0}$ ($>$$\Delta_{0}$) the states have
undergone a transition and they are deconfined asymptotically with
an oscillatory tail. This behaviour of the states with respect to
the potential strength is valid for all values of $m$ and
$\tau$=$\pm$1; though, as shown in Fig.~\ref{qstates}, the
relative amplitude of the two components depends on the choice of
$\tau$. In particular, for $\tau$=+1, $f_{1}$$>$$f_{2}$
($f_{1}$$<$$f_{2}$) when $E$$>$$0$ $(E$$<$$0)$, whereas the
opposite is valid for $\tau$=$-$1, $f_{1}$$<$$f_{2}$
($f_{1}$$>$$f_{2}$) when $E$$>$$0$ $(E$$<$$0)$.~\cite{note2}

A similar confinement-deconfinement transition in the character of
the dot states can also be induced for $\Delta$=0, provided that
the vector potential is nonzero ($A_{\theta}$$\neq$0) and has the
same power-law dependence as the electrostatic
potential.~\cite{giavaras2009} This can happen because the first
and third terms in Eq.~(\ref{k2dependence}) have the same sign.
This observation might be the key to fabricate a graphene dot with
the help of a uniform magnetic field and an electrostatic
potential, and some preliminary calculations suggest that this
should be feasible in realistic dot designs formed in gated
graphene.~\cite{giavaras2009,giavaras2009b}

\begin{figure}
\includegraphics[height=8.0cm]{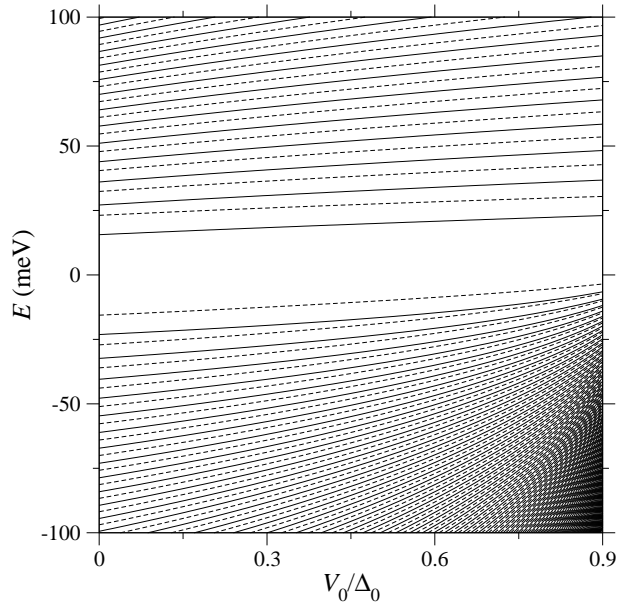}\\
\caption{Energy levels of confined states for the two valleys:
$\tau$=$+$1 (solid lines) and $\tau$=$-$1 (dashed lines). The
angular momentum number is $m$=2, the electrostatic potential is
modelled by $V$=$V_{0}r^{2}$ and the mass term is modelled by
$\Delta$=$\Delta_{0}r^{2}$. Here $\Delta_{0}$=1 $\mu$eV nm$^{-2}$,
whereas $V_{0}$ is tuned.}\label{modellevels}
\end{figure}

Figure~\ref{modellevels} shows the energy level diagram in a range
of confined states for the two valleys $\tau$=$\pm$1. The spectrum
consists of two ladders (sets) of discrete levels indicating the
existence of confined states. The energy ladders are separated by
a gap which is large when the angular momentum is large. As
$V_{0}$ increases the confinement for negative energies becomes
weak and for this reason the splitting between the discrete levels
cannot be clearly resolved. The pattern of the two ladders is
characteristic of confined states in graphene quantum dots and,
for instance, can also occur when the mass term is replaced by a
vector potential of the same spatial
dependence.~\cite{giavaras2009b}

\section{Investigation of a Dirac gap-induced dot}\label{basic}

In this section the basic properties of a Dirac gap-induced dot
are investigated numerically. We assume that there is neither an
electrostatic potential ($V$$=$0) nor a magnetic field
($A_{\theta}$$=$0) and thus confinement is due solely to the
spatial modulation of the gap. Then confinement depends on energy,
as can be seen from Eq.~(\ref{k2dependence}), and it is achieved
when $E^2-\Delta^2$$<$0, where $\Delta$ is the value of the mass
term asymptotically.

\begin{figure}
\includegraphics[height=7.cm]{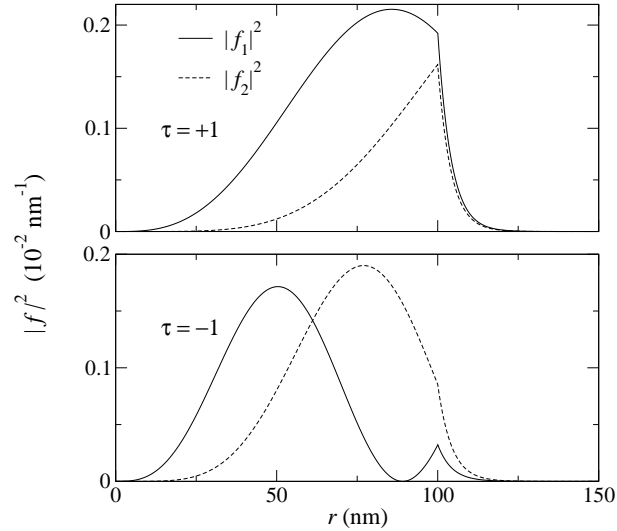}
\caption{Quantum states for a piecewise-constant mass term
$\Delta$ given by Eq.~(\ref{piecewise}). The parameters are $m$=2,
$R$=100 nm, and $\delta_{0}$=80 meV. Results for both valleys,
$\tau$=$\pm$1, are shown. The states correspond to the lowest
positive eigenenergy.}\label{analyticstates}
\end{figure}

\subsection{Graphene dot formed by a piecewise-constant Dirac gap}

First a regime for which analytical results can be obtained is
examined. Consider the special limit where the spatially-dependent
mass term $\Delta$ changes discontinuously. In this case $\Delta$
can be modelled by the expression
\begin{equation}\label{piecewise}
\Delta(r)=\delta_{0}\Theta(r-R),
\end{equation}
with $\Theta$ the Heaviside step-function. For simplicity we
assume that $\Delta(r)$=0 for $r$$<$$R$, although in the following
analysis a nonzero value can be introduced straightforwardly. As
shown in the previous section, the two radial components $f_{1}$
and $f_{2}$ satisfy a second-order differential equation, which
for $V$=0 and $A_{\theta}$=0 has the form
[$(\tau\Delta)^{2}$=$\Delta^{2}$]
\begin{subequations}
\begin{eqnarray}
\frac{d^2f_{1}}{dr^2}-\frac{4(m-1)^2-1}{4r^2}f_{1}+\frac{E^2-\Delta^2}{\gamma^2}f_{1}=0,\label{deq1}
\\
\frac{d^2f_{2}}{dr^2}-\frac{4m^2-1}{4r^2}f_{2}+\frac{E^2-\Delta^2}{\gamma^2}f_{2}=0.\label{deq2}
\end{eqnarray}
\end{subequations}
If we assume a solution of the form
$f_{i}$=$\sqrt{r}\mathcal{F}_{i}$, then $\mathcal{F}_{i}$ have to
satisfy the following Bessel's differential equations
\begin{subequations}
\begin{eqnarray}
\frac{d^2\mathcal{F}_{1}}{dr^2}+\frac{1}{r}\frac{d\mathcal{F}_{1}}{dr}-\frac{(m-1)^2}{r^2}\mathcal{F}_{1}+\frac{E^2-\Delta^2}{\gamma^2}\mathcal{F}_{1}=0,
\\
\frac{d^2\mathcal{F}_{2}}{dr^2}+\frac{1}{r}\frac{d\mathcal{F}_{2}}{dr}-\frac{m^2}{r^2}\mathcal{F}_{2}+\frac{E^2-\Delta^2}{\gamma^2}\mathcal{F}_{2}=0.
\end{eqnarray}
\end{subequations}
Setting $k$=$E$/$\gamma$ and
$q$=$\sqrt{|E^2-\delta_{0}^{2}|}$/$\gamma$, the solutions for
confined states can be written in the general form
\begin{equation}\label{rlr0}
\left(%
\begin{array}{c}
  \mathcal{F}_{1} \\
  \mathcal{F}_{2} \\
\end{array}%
\right)=\left(%
\begin{array}{c}
 \alpha J_{m-1}(k r) \\
 \alpha J_{m}(k r) \\
\end{array}%
\right), \qquad     r\leq R,
\end{equation}
and
\begin{equation}\label{rgr0}
\left(%
\begin{array}{c}
  \mathcal{F}_{1} \\
  \mathcal{F}_{2} \\
\end{array}%
\right)=\left(%
\begin{array}{c}
 \beta K_{m-1}(q r) \\
 c \beta  K_{m}(q r) \\
\end{array}%
\right), \qquad    R\leq r,
\end{equation}
with $c$=$\sqrt{|E^2-\delta_{0}^2|}/(E+\tau\delta_{0})$ and
$\alpha$, $\beta$ are constants that can be determined from the
normalisation condition and the requirement that both components
are continuous at $r$=$R$. Here $J_{m}$ is an ordinary Bessel
function of the first kind, and $K_{m}$ is a modified Bessel
function of the second kind. These have been chosen since $J_{m}$
is regular at the origin ($r$=0), while $K_{m}$ decays
exponentially at large radial distances as needed for confined
states.~\cite{boas} The coefficients ($\alpha$, $\beta$, $c$) in
Eqs.~(\ref{rlr0}) and~(\ref{rgr0}) are introduced in order $f_{1}$
and $f_{2}$ to satisfy the coupled Eqs.~(\ref{radialequations1})
and~(\ref{radialequations2}).

Both components have to be continuous at $r$=$R$, leading to the
following algebraic equation
\begin{equation}\label{bessel}
cJ_{m-1}(k R)K_{m}(q R)=J_{m}(k R)K_{m-1}(q R).
\end{equation}
This is solved numerically, with bisection, to give the energies
of the confined states in the range $|E|$$<$$\delta_0$. Thanks to
the properties of the Bessel functions: $J_{-m}$=$(-1)^{m}J_{m}$
and $K_{-m}$=$K_{m}$, the energies satisfy the condition
$E(m,\tau)$=$E(1-m,-\tau)$ as time-reversal symmetry requires when
$A_{\theta}$$=$0. Figure~\ref{analyticstates} shows one example of
the two confined components for the lowest positive energy. When
this is much smaller than $\delta_{0}$ the states decay quickly in
the region $r$$>$$R$, while the states exhibit a kink at $r$=$R$
because their first derivative are discontinuous at that point.

\subsection{Graphene dot formed by a continuous Dirac gap profile}

In any realistic graphene system the spatial modulation of the
Dirac gap will not be perfectly sharp. For this reason we make a
more realistic choice of the mass term, than that in
Eq.~(\ref{piecewise}), and use the expression
\begin{equation}\label{massterm}
\Delta(r)=\left\{
\begin{matrix}
0,            \qquad     r \leq R \\ \\
-\delta_{0}\cosh\left( \frac{r-R}{d}\right)^{-2}+\delta_{0},
\qquad R \leq r,
\end{matrix}
\right.
\end{equation}
so that asymptotically $\Delta$$\approx$$\delta_0$. For all the
calculations we choose for the parameters $R$=250 nm and $d$=150
nm, although these choices do not affect the main conclusions. In
the limit $d$$\rightarrow$0 we recover Eq.~(\ref{piecewise}).

\begin{figure}
\includegraphics[height=6.7cm]{fig4a_prb}
\includegraphics[height=6.7cm]{fig4b_prb}
\includegraphics[height=7.3cm]{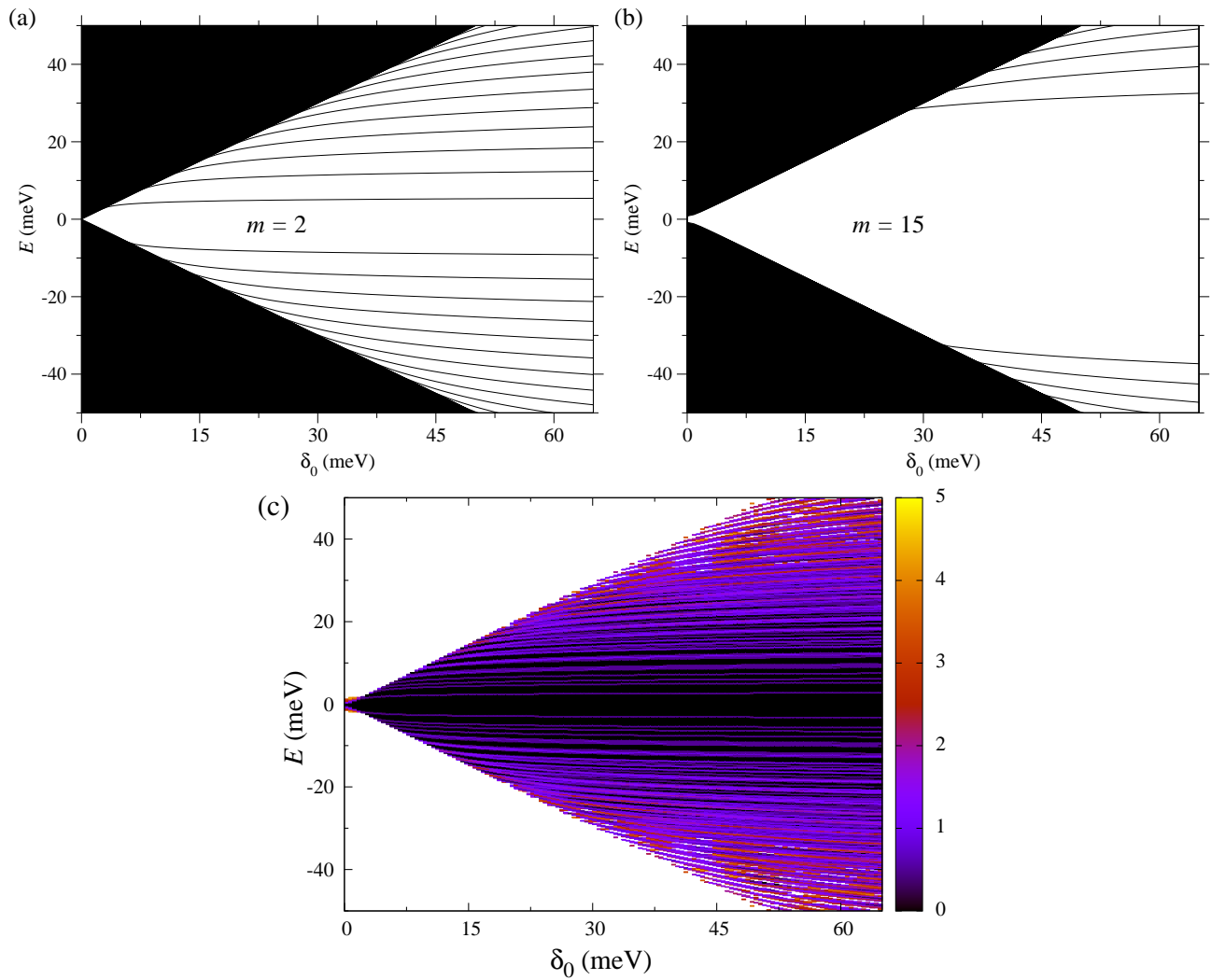}\\
\caption{(Color online) (a), (b) Energy levels, for $\tau$=$+1$,
as a function of the asymptotic value of the mass term
$\delta_{0}$ which generates a Dirac gap of $2\delta_{0}$. The
mass term $\Delta$ is modelled by Eq.~(\ref{massterm}). (c) The
contour plot shows the density of states; the maximum number of
states is restricted to five.}\label{vbzero}
\end{figure}

Figure~\ref{vbzero} shows the energy levels as a function of the
asymptotic value of the mass term $\delta_0$, giving rise to an
energy gap of $2\delta_{0}$. The (quasi) continuum of levels,
corresponding to the black area in Figs.~\ref{vbzero}(a) and (b),
reflect deconfined states. The finite system size needed for the
numerical calculations generates a gap at $\delta_{0}$=0, which
increases with $|m|$. However, the system size is chosen large
enough and hence it does not affect the physics of the discrete
levels of the dot that we are interested in. These levels emerging
through the continuum reflect the formation of confined states.
The transition from continuum to discrete levels, and vice-versa,
is accompanied by the appearance of anticrossing
points.~\cite{giavaras2010b} As seen in Figs.~\ref{vbzero}(a) and
(b), the discrete levels form two ladders (sets) of energy with
either positive or negative values, separated by a gap pertinent
to the angular motion. The typical splitting of the levels
increases with $\delta_{0}$ because the confinement becomes
stronger. Further, the two energy ladders are formed after a
critical value of $\delta_0$ depending on the angular momentum.
This happens because the angular motion tends to delocalize the
states near the origin of the dot, see Eq.~(\ref{uterm}), and
therefore for a given $\delta_0$ not all $m$ values produce
confined states. For this reason there are no discrete levels for
$\delta_{0}$$\lesssim$3 meV and 28 meV in Fig.~\ref{vbzero}, when
$m$=2 and $m$=15, respectively.

If $|m|$ is large, then $\delta_0$ has to be large for the
formation of confined states. This effect has a direct signature
in the density of states (DOS) which is shown in the contour plot
of Fig.~\ref{vbzero}(c). For simplicity only $\tau$=+1 is
considered, since inclusion of $\tau$=$-$1 simply doubles the
number of states. Moreover, because $V$=$A_{\theta}$=0 the
energies satisfy the condition $E(m,\tau)$=$-E(1-m,\tau)$ and thus
the DOS is symmetric with respect to $E\rightarrow-E$. For a small
$\delta_0$ only a few energy levels near zero lie within the gap,
corresponding to small angular momentum states. The general trend
is that with increasing $\delta_0$ the number of discrete levels
that falls in the gap increases since gradually larger $m$ values
give confined states. As an example, for $\delta_{0}$=5 meV the
discrete levels correspond to angular momentum numbers $-$1
$\leqslant$$m$$\leqslant$ 2, whereas for $\delta_{0}$=20 meV to
$-$9 $\leqslant$$m$$\leqslant$ 10. Still, however, for energies
near the middle of the Dirac gap, $-$5 meV$\lesssim$$E$$\lesssim$
5 meV, the DOS remains low regardless of $\delta_{0}$, because
only small $m$ values contribute to this energy range. This
suggests that the formation of a gap-induced dot can be probed
using similar charge measurements as, for example, in GaAs dots.
The Fermi energy of the graphene dot has to be adjusted near the
middle of the energy gap where the DOS is expected to be low and
scanning tunneling microscopy could be used to probe the confined
quantum states.

\begin{figure}
\begin{centering}
\includegraphics[width=8.2cm]{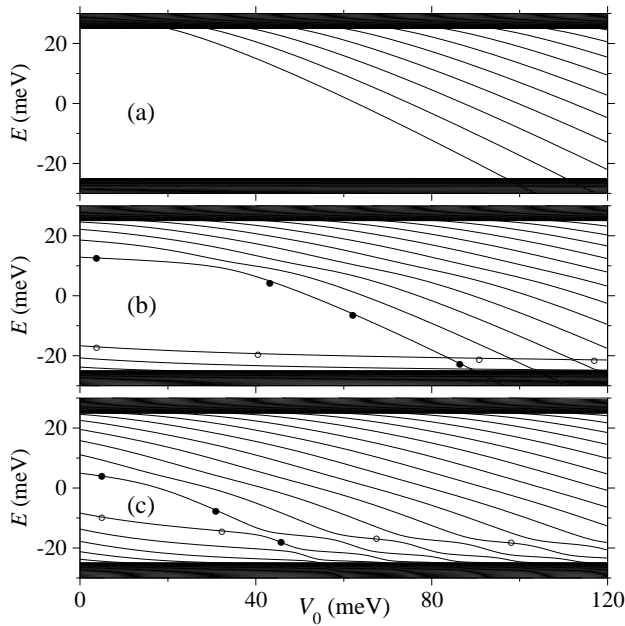}\\
\caption{(a) Energy levels versus the electrostatic potential
depth $V_{0}$, for $m$=6, $\tau$=+1 and a constant Dirac gap of
$2\Delta$=50 meV. (b) As in (a) but for a spatially modulated
Dirac gap with an asymptotic value of $2\delta_{0}$=50 meV. (c) As
in (b) but for $m$=2. The states of the energies marked by circles
are shown in Fig.~\ref{states}.}\label{gap}
\end{centering}
\end{figure}

\section{Effect of a quantum well potential on a gap-induced
dot}\label{gappotential}

In this section the effect of an electrostatic quantum well
potential on a gap-induced dot is examined. It is assumed that the
potential can be generated by a gate electrode and is not related
to the presence of the Dirac gap. We consider the most common
experimental regime where both the electrostatic potential $V$ and
the mass term $\Delta$ are constant asymptotically. In this case,
if $A_{\theta}$=0 then confinement depends on energy, as can be
seen from Eq.~(\ref{k2dependence}), and it is achieved when
$(V-E)^2-\Delta^2$$<$0. If $A_{\theta}$ has a power-law dependence
the states are always confined independent of their energy, but in
this work we focus on the case where $A_{\theta}$=0, and therefore
there is no magnetic field in the graphene sample.

The interplay of a quantum well potential with the gap-induced dot
has a drastic effect on the resulting states. In particular,
gap-induced dot states couple to the states induced by the
potential. As a result the region in which the resulting
hybridized states are localised is tunable with the strength of
the potential. Also, Klein tunneling in the electrostatic barrier
region occurs.

It has been experimentally demonstrated that gate electrodes can
be suspended above the graphene sheet inducing a smooth quantum
well or barrier potential that is tunable with the applied gate
voltage.~\cite{velasco,gorbachev} The exact potential profile can
be determined by solving the Poisson equation within a
semiclassical Thomas-Fermi model properly taking into account
charge screening effects and the specific electrode
geometry.~\cite{gorbachev,giavaras2009} However, a smooth and
slowly varying potential can be modelled, to a good approximation,
by the Gaussian expression~\cite{giavaras2009,giavaras2009b}
\begin{equation}
V(r)=-V_{0}\exp\left(-\frac{r^2}{l_0^{2}}\right).
\end{equation}
The parameters $V_{0}$ and $l_{0}$ model the effective depth and
width of the quantum well potential, which are controlled by the
geometry of, and the applied voltage to the gate electrodes.
Typical gate voltages of a few volts generate a potential well
depth of some hundreds of meV, with an effective width of some
hundreds of nm. This work is concerned only with the regime where
the quantum well potential is formed inside the spatial region
where $\Delta$$\lesssim$$\delta_{0}$, with $\Delta$ as in
Eq.~(\ref{massterm}), therefore for all the calculations
$l_{0}$=180 nm and $V_{0}\leqslant$ 120 meV.

The energy level diagram as a function of $V_{0}$ is shown in
Fig.~\ref{gap}(a), for a spatially-independent (constant) Dirac
gap with $2\Delta$=50 meV. Confined states have discrete energy
levels that lie within the gap, whereas deconfined states form an
upper and a lower band of (quasi) continuum of levels, for
energies $E$$>$$\Delta$ and $E$$<$$-\Delta$, respectively. If
$V_{0}$ is small, the angular momentum delocalizes the states;
therefore confined states are formed after a critical value of
$V_{0}$, that is $\sim$19 meV in Fig.~\ref{gap}(a). As $V_{0}$
increases, the number of discrete levels increases due to the
stronger confinement. Energy levels emerge from the upper
continuum into the gap, while the lowest discrete levels merge
into the lower continuum via anticrossing points. These are formed
inside the two bands as shown in Ref.~\onlinecite{giavaras2010b}.
The quantum states undergo a transition from confined to
deconfined and vice-versa. Specifically, a confined state transits
to a deconfined via tunneling into the valence (lower) band, when
its energy falls below the gap.~\cite{steele} In this regime the
deconfined state has an oscillatory tail but has a large amplitude
near the origin of the dot (see below).

\begin{figure}
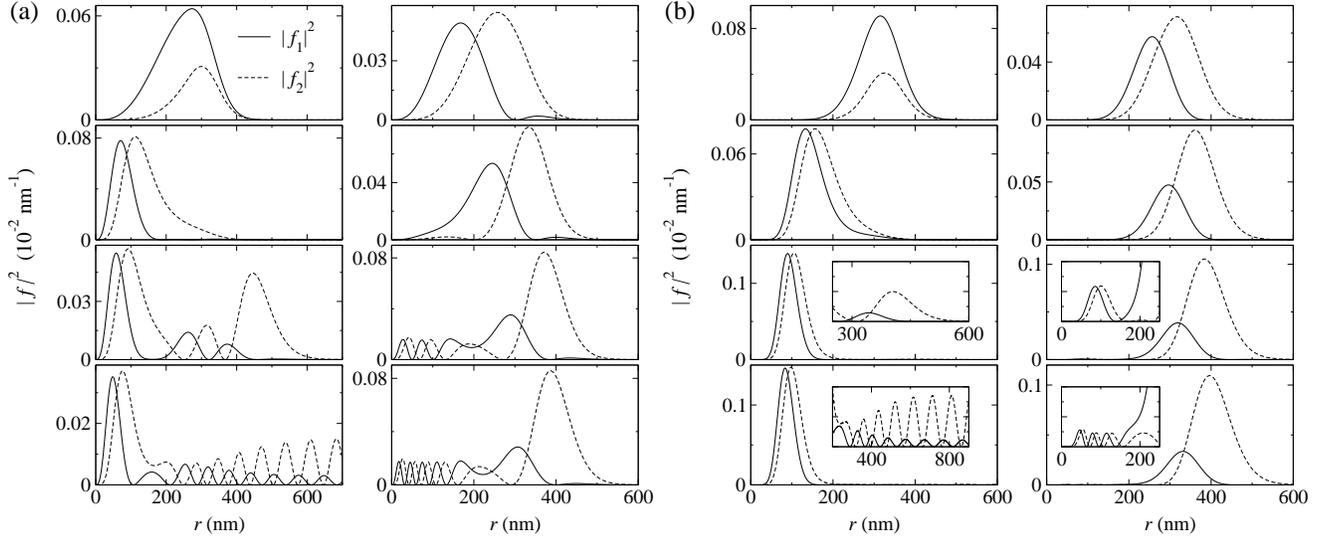

\includegraphics[width=8.6cm]{fig6a_prb}
\includegraphics[width=8.6cm]{fig6b_prb}\\
\caption{(a) Quantum states for $m$=2 and for different
electrostatic potential depths $V_{0}$; from top to bottom:
$V_{0}$=0, 30, 50, 70 meV for the left panels, and $V_{0}$=0, 30,
70, 100 meV for the right panels. The Dirac gap is spatially
modulated with an asymptotic value of $2\delta_{0}$=50 meV. Left
(right) panels show states with energies marked by $\bullet$
($\circ$) in Fig.~\ref{gap}(c). (b) As in (a) but for $m$=6 and
$V_{0}$=0, 40, 85, 100 meV for the left panels, and $V_{0}$=0, 40,
85, 120 meV for the right panels. Left (right) panels show states
with energies marked by $\bullet$ ($\circ$) in Fig.~\ref{gap}(b).
The vertical axes of the insets range from 0 to
1$\times$10$^{-3}$.}\label{states}
\end{figure}

Figures~\ref{gap}(b) and (c) show the energy diagram when the
Dirac gap is spatially modulated with an asymptotic value of
$2\delta_{0}$=50 meV. The appearance of discrete levels for
$V_{0}$=0 is a consequence of the gap-induced confinement. Such
levels are absent in the constant gap system. Furthermore, for
$V_{0}$=0 there are more discrete levels for $m$=2 than $m$=6, as
expected based on the above arguments. Anticrossing points, formed
now between discrete levels, indicate a coupling between
potential- and gap-induced states. The gap-induced states of the
upper energy ladder couple strongly to the potential states and
for this reason the corresponding anticrossing points are not
well-formed. On the other hand, states of the lower energy ladder
couple weakly to the electrostatic potential for $m$=6
[Fig.~\ref{gap}(b)], and therefore the anticrossing points are
well-formed, while this coupling is much stronger for $m$=2
[Fig.~\ref{gap}(c)].

Figure~\ref{states} illustrates the dot states as the potential
depth $V_{0}$ increases. Consider first the $m$=2 states shown in
the left panels of Fig.~\ref{states}(a). Confinement for $V_{0}$=0
is due to the spatial modulation of the Dirac gap and cannot be
realised for a constant gap. As $V_{0}$ increases, the gap-induced
state couples to the lowest energy state induced by the potential
well ($V_{0}$=30 meV), and its energy decreases. The state
acquires a large amplitude in the potential well region and with
increasing $V_{0}$ it couples successively to gap-induced states
of the lower energy ladder ($V_{0}$=50 meV). In this range the
resulting hybridised state is spread with significant amplitude
over the whole non-asymptotic region defined by
$\Delta\lesssim\delta_{0}$. Further increasing $V_{0}$, the energy
of the state falls below the gap ($V_{0}$=70 meV); the state
tunnels in the valence band and becomes deconfined with an
oscillatory tail.

Strong coupling between gap- and potential-induced states occurs
also for the $m$=2 states shown in the right panels of
Fig.~\ref{states}(a); for instance, for $V_{0}$=70 meV and 100
meV. With increasing $V_{0}$, the gap-induced state that
corresponds to the highest energy in the lower ladder (for
$V_{0}$=0) couples successively to excited potential-induced
states. This effect is reflected in the energy level diagram
[Fig.~\ref{gap}(c)] by the appearance of a series of anticrossing
points as $V_{0}$ increases. A similar effect is displayed by
gap-induced states of smaller energy, although with increasing
$V_{0}$ the number of gap-induced confined states in the lower
energy ladder decreases. For instance, as seen in
Fig.~\ref{gap}(c) for $V_{0}$=0 there are five discrete levels,
whereas for $V_{0}$=50 meV there are three. The number of states
can be determined using a similar approximate formalism as that in
Ref.~\onlinecite{giavaras2010a}. The necessary condition for the
existence of confined states is that there must be a spatial
region near $r$=$R$ in which $k^{2}_{2}$$>$0, with $k^{2}_{2}$ as
given in Eq.~(\ref{k2exact}), whereas asymptotically
$k^{2}_{2}$$<$0.

In Fig.~\ref{states}(a) the coupling between potential- and
gap-induced states of the lower ladder is strong and therefore the
resulting hybridised states have a relatively strong amplitude
over the whole region where $\Delta\lesssim\delta_{0}$. In
contrast, the coupling is weak for the $m$=6 states shown in
Fig.~\ref{states}(b). Therefore, for $V_{0}$=85 meV, the states
peak mainly in the region defined either by the potential well
(left panel) or the Dirac gap modulation (right panel). These two
regions have a small overlap when $m$ is large and the effective
width of the potential well is smaller than the radius of the
zero-gap region.

The $m$-dependent coupling of the states is related to the width
of the classically forbidden region formed between the confinement
regions defined by the gap modulation and the electrostatic well.
The semiclassical approach developed in Ref.~\onlinecite{chen}
shows that for a large positive $m$ the width of the forbidden
region is large. This results in a weak coupling between the
potential- and gap-induced states of the lower energy ladder.
Also, as $m$ increases the gap between the upper and the lower
energy ladders increases too, (Fig.~\ref{vbzero}) and therefore
$V_{0}$ has to be large for the formation of coupled states. The
coupling effect, though, is insensitive to the details of the gap
modulation. Our calculations show that both weak and strong
couplings can be realised even when the Dirac gap changes
discontinuously as in Eq.~(\ref{piecewise}).

\begin{figure}
%Requires \usepackage{graphicx}
\includegraphics[height=6.8cm]{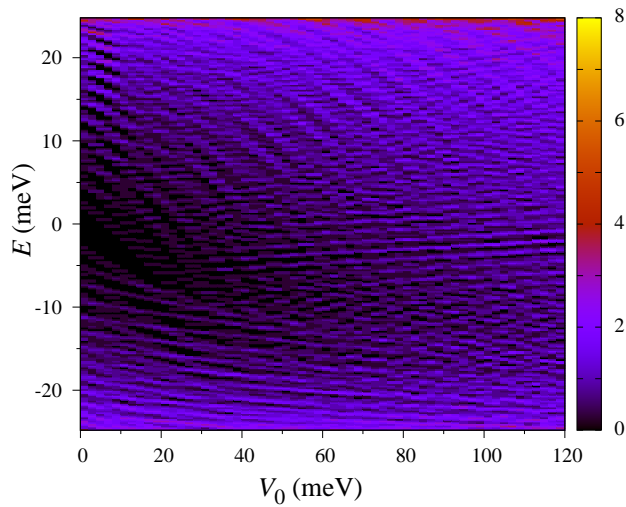}
\caption{(Color online) Density of states versus energy $E$
($|E|<$ 24.8 meV) and electrostatic potential depth $V_{0}$, for
$\tau$=+1, and an asymptotic Dirac gap of $2\delta_{0}$=50
meV.}\label{v0dos}
\end{figure}

When the potential is nonzero ($V_{0}$$\neq$0) the induced states
can have an oscillatory amplitude inside the barrier region,
though asymptotically they decay.~\cite{note3} This effect is a
consequence of the Klein tunneling in the non-asymptotic region
defined by $\Delta\lesssim\delta_{0}$. In contrast, the quantum
states of a Schr\"odinger dot display only an exponential decay
inside an electrostatic barrier. Numerical calculations of the
DOS, shown in Fig.~\ref{v0dos}, suggest that it should be
experimentally possible to resolve graphene states with large
oscillatory amplitude in the barrier and the coupling between the
states described above. In order to probe the states with scanning
tunneling microscopy the DOS has to be low. This could be achieved
for a small $V_{0}$, since in this case only energies that
correspond to small $m$ states fall in the gap. For the same
reason the DOS is low when the asymptotic value of the Dirac gap
is small.

\section{Discussion and conclusions}\label{conclusions}

The existence of a Dirac gap in the energy spectrum of monolayer
graphene suppresses the Klein tunneling and thus enables charge
confinement by an electrostatic quantum well and formation of
quantum dots. Here it was shown that a graphene dot can be formed
as a result of a spatially modulated Dirac gap. It was found, by
solving the Dirac equation in the continuum limit, that when the
gap has a local minimum confined states with discrete energy
levels can be formed, without applying external electric and/or
magnetic fields. This cannot be achieved in a constant Dirac gap
graphene system. The required gap modulation may be introduced
with substrate engineering, local strain, or a chemical technique.
Unlike quantum dots formed in nanocrystals of graphene, the
proposed dot is formed in a large sheet of graphene, therefore the
physics of the edges is unimportant and does not affect the dot
properties.

The general conditions for confinement in the presence of a
spatially modulated Dirac gap, an external electrostatic
potential, and a magnetic vector potential were analysed. When the
gap and/or the vector potential rises asymptotically, while the
electrostatic potential is zero (constant), the resulting states
are confined regardless of angular momentum, valley and
eigenenergy. The interplay of the gap-induced dot with an
electrostatic potential leads to tunable quantum states, i.e.,
from confined to deconfined and vice-versa. This can only happen
provided the potential and the mass term, which generates the
Dirac gap, have the same spatial forms asymptotically, while the
magnetic field is zero. In this case the states can be tuned with
the strength of the potential: the states are deconfined when the
electrostatic potential is stronger than the mass term and
confined in the opposite regime.

When the Dirac gap is zero within a disc area and constant outside
that area confinement is energy-dependent, thus the choice of
angular momentum and valley is important. Confined states are
localised inside the disc area when their energies lie in the gap,
otherwise the states are deconfined. The energy spectrum of the
confined states consists of two ladders (sets) of discrete levels
separated by a gap. Numerical calculations of the DOS suggest that
states with small angular momentum lie in a region of low density
and hence they could be probed using standard techniques such as
scanning tunneling microscopy.

It was also shown that states induced by a quantum well potential,
which may be generated by a gate electrode, can coexist and couple
to gap-induced dot states. When the coupling is weak the states
retain their character, whereas in the opposite regime the states
are strongly hybridised. The signature of this coupling in the
energy spectrum is the appearance of a series of anticrossing
points. This coupling property offers a way of tuning the spatial
region, in which the hybridised states are localized, with the
strength of the potential. Moreover, the states can have a large
oscillatory amplitude in the barrier region exhibiting Klein
tunneling, before they eventually decay asymptotically.
Calculations of the DOS indicate that these quasi-relativistic
effects could be probed.

\section*{ACKNOWLEDGEMENTS}

We thank P.A.~Maksym for comments on the manuscript. G.G.
acknowledges support from the Japan Society for the Promotion of
Science (JSPS). F.N. acknowledges partial support from the
Laboratory of Physical Sciences, National Security Agency, Army
Research Office, AFOSR, DARPA, National Science Foundation grant
No.~0726909, JSPS-RFBR contract No.~09-02-92114, Grant-in-Aid for
Scientific Research (S), MEXT Kakenhi on Quantum Cybernetics, and
Funding Program for Innovative R$\&$D on S$\&$T (FIRST).


\begin{thebibliography}{10}
\bibitem{nanoflakes}
Quantum dots formed in nanoflakes of graphene have been fabricated
(see Ref.~\onlinecite{ponomarenko}), however this work is
concerned with dots formed in a large sheet of graphene, where the
edges of the sheet have no effect on the dots properties and can
be ignored.
\bibitem{note0}
The symbol $E(m, B, \tau)$ denotes the set of eigenergies given by
Eqs.~(\ref{radialequations1}) and~(\ref{radialequations2}) for
angular momentum number $m$, magnetic field $B$ and valley $\tau$.
The radial quantum number is not stated explicitly.
\bibitem{note1}
The component $f_{1}$ satisfies a similar Schr\"odinger-like
equation as $f_{2}$ but with $k^{2}_{1}$$\neq$$k^{2}_{2}$.
However, asymptotically $k^{2}_{1}$=$k^{2}_{2}$.
\bibitem{note2}
We are not concerned with a possible relative difference between
the two components $f_{1}$ and $f_{2}$, but only with the relative
maximum amplitude of the two components.
\end{thebibliography}
\end{document}